\documentclass[aps, prl, reprint, superscriptaddress]{revtex4-1}
\usepackage{graphicx}
\usepackage[colorinlistoftodos]{todonotes}
\usepackage[normalem]{ulem}
\usepackage{xcolor}
\usepackage{hyperref}
\begin{document}

\title {Negative thermal expansion in the plateau state of a 
magnetically{--}frustrated spinel}

\author{L. Rossi}
\author{A. Bobel}
\author{S. Wiedmann}
\affiliation{High Field Magnet Laboratory (HFML-EMFL), Radboud University, Nijmegen, Netherlands}
\affiliation{Radboud University, Institute of Molecules and Materials, Nijmegen, Netherlands}
\author{R. K\"uchler}
\affiliation{Max Planck Institute for Chemical Physics of Solids, N{\"o}thnitzer Str. 40, 01187 Dresden, Germany}
\author{Y.~Motome}
\affiliation{Department of Applied Physics, The University of Tokyo, Hongo 7-3-1, Bunkyo-ku, Tokyo 113-8656, Japan}
\author{K. Penc}
\affiliation{Institute for Solid State Physics and Optics, Wigner Research Centre for Physics, Hungarian Academy of Sciences, Budapest, Hungary}
\author{N. Shannon}
\affiliation{Okinawa Institute of Science and Technology Graduate University, 
{Onna--son, Okinawa 904--0495}, Japan}
\affiliation{{Department of Physics, Technische Universit\"at M\"unchen, D-85748 Garching, Germany}}
\author{H. Ueda}
\affiliation{Department of Chemistry, Graduate School of Science, Kyoto University, Kyoto, Japan
}
\author{B. Bryant}
\affiliation{High Field Magnet Laboratory (HFML-EMFL), Radboud University, Nijmegen, Netherlands}
\affiliation{Radboud University, Institute of Molecules and Materials, Nijmegen, Netherlands}

\date{\today}

\begin{abstract}
We report on negative thermal expansion (NTE) in the high--field, half--magnetization plateau phase of the frustrated magnetic insulator CdCr$_{2}$O$_4$. Using dilatometry, we precisely map the phase diagram at fields of up to $30\ \text{T}$, and identify a strong NTE associated with the collinear half--magnetization plateau for $B > 27\ \text{T}$. The resulting phase diagram is compared with a microscopic theory for spin--lattice coupling, and the origin of the NTE is identified as a large negative change in magnetization with temperature, coming from a nearly--localised band of spin excitations in the plateau phase. These results provide useful guidelines for the discovery of new NTE materials.
\end{abstract}

\maketitle

Frustrated magnets are materials with competing spin interactions, which cannot be simultaneously satisfied. 
While these materials are most famous as  a playground for novel phases such as quantum spin liquids \cite{Lee2008,Savary2016}, they also exhibit other, more technologically--relevant properties, such as multiferroicity \cite{Cheong2007,Tokura2014,Fiebig2016}, and an enhanced magnetocaloric effect \cite{Zhitomirsky2003,Zhitomirsky2004,Zhitomirsky2005,Derzhko2006,Schmidt2007}. 
Negative thermal expansion (NTE) is another unusual phenomenon often observed in frustrated magnets \cite{Ramirez2000, Shiga1993, Hemberger2007, Li2016}. This effect provides a route for the control of thermal expansion necessary to ensure the performance of high--precision devices \cite{Takenaka2018}, so theoretical models which can act as a guide for discovery of new NTE materials are highly valuable.

In frustrated magnets with a strong coupling between the spin and lattice degrees of freedom, the interplay between magnetic field and spin--lattice coupling produces a range of phases in which frustration is partially relieved, an effect known as ``order by distortion'' \cite{Yamashita2000,Tchernyshyov2002PRL,Tchernyshyov2002PRB,Penc2004, Motome2006, Bergman2006,Shannon2010}. A paradigm for this type of behaviour is provided by Cr--spinels, which exhibit many different magnetically--ordered phases as a function of magnetic field \cite{Ueda2006, Rudolf2007, Kojima2008, Tsurkan2011, Miyata2011-JPSJ80,Miyata2011-PRL107,Miyata2012}. Many of these systems exhibit NTE \cite{Hemberger2007, Yokaichiya2009,Tachibana2011}, including the spinel CdCr$_{2}$O$_4$ in zero magnetic field \cite{kitani2013thermal}. This suggests that the unusual thermodynamic behaviour may have a common origin; however to date there is no general understanding of this phenomenon, or how it is linked to spin--lattice coupling. Moreover, to obtain a complete picture of NTE in spinels, high--precision measurements are also needed for the ordered phases induced by magnetic field.

In this Letter, we report on thermal expansion and magnetostriction measurements of the frustrated spinel CdCr$_{2}$O$_4$, in magnetic fields up to 30~T. We map the phase diagram, which we compare to that derived from a microscopic model of spin--lattice coupling. The high--field half--magnetization plateau phase exhibits enhanced thermal stability compared to theory, characteristic of a strong spin--lattice coupling in this phase. This state also shows a marked NTE, distinct from that observed in zero field. Starting from the same model of spin--lattice coupling, we develop a microscopic theory of this NTE, and identify its origin as being a band of nearly--localised magnetic excitations. These results provide a general framework for modelling and predicting NTE in pyrochlore lattices, and in frustrated magnets in general.

The pyrochlore lattice, which consists of corner sharing tetrahedra, is a well--known stage for strong geometric frustration \cite{Gardner2010}. This structure is realized in the position of the Cr$^{3+}$ ions in the chromium spinels ACr$_{2}$X$_4$, where A is Zn, Cd or Hg and X is O, S or Se. The strength and sign of the Cr-Cr spin coupling depends strongly on the interatomic distance \cite{yaresko2008,ueda2008pressure}, leading to a strong coupling between spin ordering and lattice distortions. The oxide spinels ACr$_{2}$O$_4$ all have antiferromagnetic spin coupling and are magnetically frustrated: because of the frustration they remain paramagnetic down to temperatures well below the Curie--Weiss temperature $\Theta_{CW}$. At T$_{N}$ the spin frustration is relieved due to a spontaneous lattice distortion \cite{Yamashita2000,Tchernyshyov2002PRL,*Tchernyshyov2002PRB}, which allows a noncollinear spin--spiral antiferromagnetic ground state \cite{matsuda2007spiral,matsuda2010}. 

\begin{figure}
	\begin{centering}
    \includegraphics[scale=1]{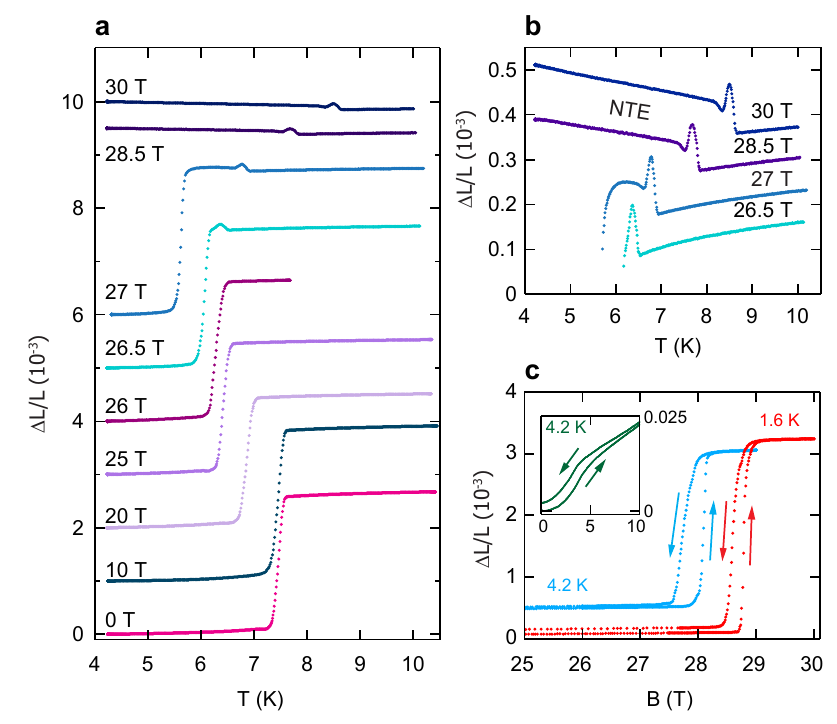}
	\caption{(color online). Thermal expansion and magnetostriction measurements of CdCr$_{2}$O$_4$ at magnetic fields up to 30~T, showing negative thermal expansion (NTE) for fields $B > 27\ \text{T}$.  These results were used to determine the phase diagram shown in Fig. \ref{Figure3}. (a) Thermal expansion measurements from 4.2~K to 10.4~K in fields up to 30~T, measured on warming, showing the variation of $T_N$ with field. (b) Detail of data in (a), for fields from 26.5~T to 30~T, showing  transition from the paramagnetic state to the half--magnetization plateau state, and the presence of NTE in the plateau state above 27~T. In (a) and (b) the curves have been offset for clarity. (c) Magnetostriction measurements from zero to 30~T at 4.2~K and 1.6~K, showing a hysteretic transition from the antiferromagnetic state to the plateau state. The inset shows magnetostriction up to 10~T at 4.2~K, showing a hysteretic low field transition at around 4.5~T.}
	\label{Figure2}
	\end{centering}
\end{figure}

The Cr oxide spinels show another magnetostructural transition at high magnetic field, into a collinear state with one--half of the saturation magnetization, in which three of the spins in each tetrahedron point ``up'', and one points ``down'' \cite{Penc2004, Ueda2005, Ueda2006, Shannon2006, Shannon2010}. This state has a constant magnetization across a wide range of magnetic fields, and it is thus often referred to as the ``plateau'' state. Both the magnetostructural transition at T$_N$ and the transition to the half--magnetization plateau are manifestations of the strong spin--lattice-coupling in the Cr oxide spinels: a developed microscopic magnetoelastic theory \cite{Penc2004,Bergman2006}, describes how the plateau state is stabilized by the spin--lattice coupling \cite{Miyata2011-JPSJ80,Miyata2012}. 

In order to probe the interplay of frustration and spin--lattice coupling, we performed thermal expansion and magnetostriction measurements of CdCr$_{2}$O$_4$ using capacitive dilatometry at low temperatures and high magnetic fields up to 30~T \cite{Kuchler2012,Kuchler2017}. This compound was chosen since it is highly frustrated, with $f=|\Theta_{CW}|/$T$_{N}\approx$ 10, high quality single crystals are available, and it is possible to reach the plateau phase in static (DC) high field facilities. So far, zero--field thermal expansion measurements \cite{kitani2013thermal}, and pulsed--field magnetostriction measurements \cite{Ueda2005}, have been reported. We measured the strain $\Delta L/L$ along the [111] direction: the magnetic field is parallel to the [111] direction. The sample is clamped between two plates in the dilatometer, thus applying a small [111] uniaxial pressure. The effect of varying the applied pressure is discussed in the Supplemental Material \footnote[1]{See Supplemental Material for additional data on the effect of varying applied stress, the derivation of Eq. (3) and discussion of the microscopic origin of $dM/dT$, including refs \cite{Chacon2015,chung2013incommensurate}\label{Supplemental}}. 
We studied a series of single--crystal samples, all of which are plate--like with wide (111) faces, around 3-5~mm in diameter and between 80~$\mu$m and 500~$\mu$m thick. 

High field measurements were carried out with the sample mounted in a compact capacitive dilatometer in a 30~T resistive Bitter magnet \cite{Kuchler2012}. Fig. \ref{Figure2}(a) presents thermal expansion from 4.2~K to 10.4~K at zero field and at field increments up to 30~T. Clearly visible up to 27~T is the magnetostructural transition at T$_{N}$, seen here on warming from the tetragonal antiferromagnetic phase to the cubic paramagnetic phase. $T_N$ decreases from 7.5~K at zero field to 5.5~K at 27~T, while the measured [111] strain at $T_N$ remains constant. Above 26.5~T, the transition from the high--temperature paramagnetic phase to the low--temperature half--magnetization plateau phase can be seen in the $\Delta L/L$ data as a peak superposed on a step, Fig. \ref{Figure2}(b). The appearance or the absence of the peak is sample dependent, while the step was present in all the measured samples. Both phases are cubic - the paramagnetic state \emph{Fd}$\bar{3}$\emph{m}, the plateau state \emph{P}$4_3 32$ \cite{matsuda2010}, so, from measurements on three samples, we can estimate a change in unit cell volume on cooling, of $ \Delta V/V \approx 2.2\pm0.9\times10^{-4}$ \footnote[2]{In this estimate we are assuming that both phases have cubic symmetry, thus neglecting any structural anisotropy induced by magnetic field. Such distortions may be regarded as second-order corrections \cite{Penc2004}}.
 We can explain this increase in volume qualitatively as part of the general principle of the magnetoelastic theory that increased magnetization leads to increased unit cell volume, if antiferromagnetic interactions are assumed \cite{Penc2004}. Below this transition and above 27~T, NTE is seen in the plateau phase, shown in Fig. \ref{Figure2}(b).

In addition to thermal expansion, constant--temperature magnetostriction measurements were made, with field sweeps up to 30~T, at temperatures between 1.3~K and 4.2~K. Fig. \ref{Figure2}(c) shows the results from 25~T to 30~T. We see a hysteretic transition from the tetragonal antiferromagnetic phase to the plateau phase, which is consistent with a first order phase transition. The sweep rate close to the transition was 0.5~T~/~min. Previous pulsed field measurements reported a colossal negative magnetostriction at the transition to the half--magnetization plateau, for both [111] and [110] directions \cite{Ueda2005,Shannon2006}. In our [111] measurements we find a positive magnetostriction at this transition \footnote[3]{The negative magnetostriction in ref. \cite{Ueda2005} might have been caused by an anomalous strain induced by thermal contraction of an adhesive used to mount the sample. H. Ueda, private communication (2019)}. This is consistent both with measurements on HgCr$_{2}$O$_4$ \cite{Tanaka2007}, and with the magnetoelastic theory \cite{Penc2004}, in which jumps in magnetization are mirrored by unit cell expansion. Both the transition with field to the plateau phase (Fig. \ref{Figure2}(c)) and the thermal transition at $T_N$ to the cubic, paramagnetic phase (Fig. \ref{Figure2}(a)) have the same sign and similar magnitude in $\Delta L/L$. This indicates that these phases have a similar unit cell, and supports the finding that the plateau phase also has overall cubic symmetry \cite{Inami2006,matsuda2010}. 

We also performed a second magnetostriction experiment in a superconductor magnet, between zero and 15~T and from 2.2~K to 7~K. The inset in Fig. \ref{Figure2}(c) presents magnetostriction data at 4.2~K, which show a hysteretic low field transition at around 4.5~T. A similar transition has previously been observed in magnetization data \cite{Kimura2006,matsuda2007spiral}. Based on ESR and optical spectroscopy measurements \cite{Kimura2006, Sawada2014} this has been interpreted as a transition from a helical structure to a commensurate canted spin structure. Neutron diffraction experiments, though, appear to rule out an incommensurate to commensurate transition \cite{matsuda2007spiral}, instead implying a rearrangement of spin spiral domains between 2.5 and 6~T. When the field is in the \emph{a-c} plane in which the spins rotate in the spiral, a spin--flop is observed: since we apply the field along the [111] direction we would expect a flop to a conical spin spiral. 

\begin{figure}[t]
    \begin{centering}
    \includegraphics[scale=1]{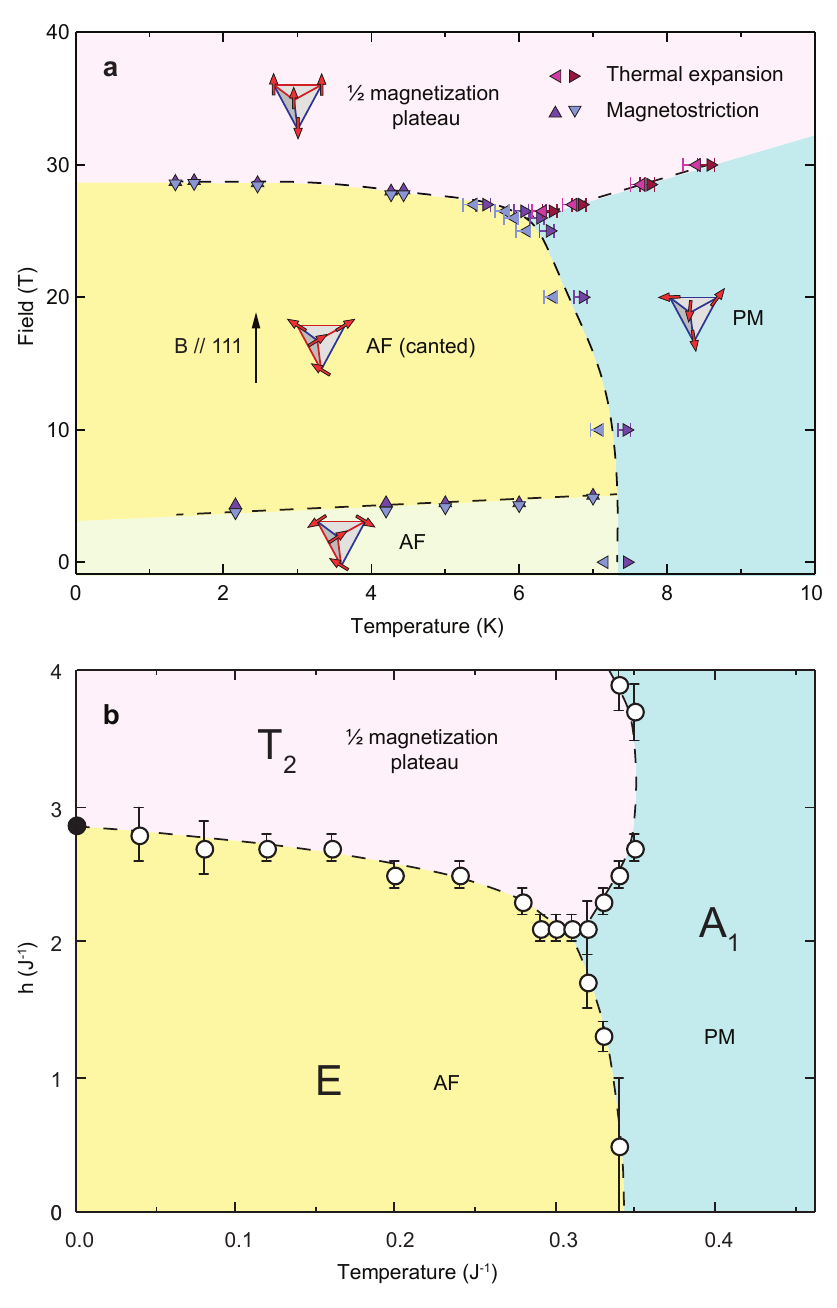}
	\caption{(color online). Low temperature, high magnetic field phase diagram for CdCr$_{2}$O$_4$. (a) Experimental phase diagram derived from magnetostriction and thermal expansion measurements, showing the antiferromagnetic (AF) and canted phases, half--magnetization plateau and paramagnetic (PM) phases. Hysteresis is shown by separate points for increasing and decreasing field or temperature. For each phase a schematic spin tetrahedron is shown, with shorter bonds in blue and longer in red. (b) Theoretical phase diagram from Monte Carlo calculations based on the magnetoelastic theory, with the spin--lattice coupling parameter $b$ = 0.1: the dominant lattice distortions in each phase ($A_1$, $E$, and $T_2$) are shown.}
	\label{Figure3}
	\end{centering}
\end{figure}

We can summarize the results from the thermal expansion and magnetostriction measurements in a phase diagram, shown in Fig. \ref{Figure3}(a). Three main phases are described: the high--temperature paramagnetic phase, the antiferromagnetic phase below 7.5~K and below 28.7~T, and the high--field half--magnetization plateau phase. Inside the antiferromagnetic phase we identify a low field transition, which increases from 4.3~T at 2.2~K to 5.1~T at 7~K. Hysteresis is observed in all the transitions. We do not find any experimental evidence of the additional phase transition recently reported from sound velocity measurements \cite{Zherlitsyn2015}, though the temperature dependence of the low field transition is consistent with that report. Our new phase diagram is more precise for fields above 12~T than previous diagrams \cite{Ueda2005,Kojima2008}.

We can use a microscopic magnetoelastic theory to reproduce the experimental phase diagram, and explain the presence of NTE in the plateau state. A simple Hamiltonian to account for the effects of spin--lattice coupling on the phase transitions in applied magnetic field in Cr spinels was introduced in \cite{Penc2004}:

\begin{equation}
 \mathcal{H} = \sum_{\langle i,j \rangle} 
  \left[
  J (1- \alpha \rho_{i,j}) \mathbf{S}_i \cdot \mathbf{S}_j
 + \frac{K}{2}  \rho_{i,j}^2 \right] -  \sum_{i} \mathbf{h}\cdot \mathbf{S}_i \;, 
 \label{eq:H}
\end{equation}
where the summation is over the nearest neighbor bonds on the pyrochlore lattice. $J$ is the antiferromagnetic exchange interaction, $\alpha$ is the spin-lattice coupling, $\rho_{i,j}$ is the change of the length of the bonds from the equilibrium distances in the paramagnetic phase, $K$ is the elastic  constant and $h$ is the applied magnetic field. In its simplest form, this theory reduces to solving an effective spin model with only two adjustable parameters, 

\begin{eqnarray}
{\mathcal H}_{\sf eff} &=& J \sum_{\langle ij \rangle} {\bf S}_i \cdot {\bf S}_j - b \left( {\bf S}_i \cdot {\bf S}_j \right)^2 
- h \sum_i S^z_i \;,
\label{eq:Heff}
\end{eqnarray}
where $b=J\alpha^2/K$ reflects the strength of the spin--lattice coupling. In the case of CdCr$_2$O$_4$, measurements of magnetization lead to an estimate of $b \approx 0.1$ \cite{Kimura2015}.

The effective spin model [Eq.~(\ref{eq:Heff})] can be solved using classical Monte Carlo calculations \cite{Motome2006, Shannon2006}, leading to the phase diagram shown in Fig.~\ref{Figure3}(b). Here, calculations have been carried out for 4--sublattice order, stabilized by an additional third--neighbor interaction $J_3 = -0.05 J$ \cite{Motome2006}. However, very similar results are obtained for 16--sublattice order \cite{Motome-unpub}. The phase diagram in Fig. \ref{Figure3}(b) has been calculated for $b=0.1$: for purposes of comparison, the results have been scaled for the experimental values of T$_N$ and the critical field H$_{c1}$.

The Monte Carlo results reproduce the experimental phases well, particularly the B-T dependence of the transitions to the antiferromagnetic phase. The main discrepancy between theory and experimental data is seen in the transition from the paramagnetic phase to the plateau phase, which experimentally has a considerably lower slope in B/T. This indicates that the plateau phase is stable to a higher temperature than the antiferromagnetic phase, as observed experimentally for HgCr$_2$O$_4$ \cite{Ueda2006}. By contrast, the Monte Carlo phase diagram (Fig. \ref{Figure3}(b)) predicts that the plateau phase is stable only up to a temperature similar to T$_N$. In a more general formulation of the magnetoelastic theory, the coefficient of spin--lattice coupling, $b$, takes on different values in phases in which tetrahedra undergo distortions with different symmetry  \cite{Penc2004,Penc2007SymmetryLattice}. In the present case, this leads to three distinct parameters;  $b_{A_1}$ (uniform changes in volume); $b_E$ (tetragonal distortions, found in the AF phase); $b_{T_2}$ (trigonal distortions, found in the half--magnetization plateau). From detailed comparison of the magnetoelastic theory to magnetization and ESR data for CdCr$_2$O$_4$, Kimura {\it et al.} \cite{Kimura2015} obtain $b_{A_1}$ = 0.05, $b_E$ = 0.1, $b_{T_2}$ = 0.14. The Monte Carlo calculations shown in Fig. \ref{Figure3}(b) assume $b_{A_1}$ = $b_E$ = $b_{T_2}$ = 0.1 so probably underestimate $b_{T_2}$, and hence the thermal stability of the plateau state, explaining the discrepancy seen between the experimental and theoretical results.

We now turn to the issue of the NTE in the plateau phase. Several spinel compounds show NTE at zero field, including CdCr$_{2}$S$_4$ \cite{Tachibana2011}, ZnCr$_{2}$Se$_4$ \cite{Hemberger2007, Chen2014}, and CdCr$_{2}$O$_4$ \cite{kitani2013thermal}. In all of these cases, the onset of NTE on cooling is in the paramagnetic phase, above the magnetic ordering temperature. Zero--field NTE in CdCr$_{2}$O$_4$ occurs exclusively within the paramagnetic phase for $45 < T < 140\ \text{K}$ \cite{kitani2013thermal}. This contrasts with the results in field, presented in Fig.~\ref{Figure2}, in which there is an abrupt onset of NTE at the magnetic ordering temperature, and the NTE occurs only within the low-temperature ordered phase. This suggests that the NTE observed within the plateau phase of CdCr$_{2}$O$_4$, may have a qualitatively different origin from that observed in the paramagnetic phase in zero field.

NTE in pyrochlore lattices is often attributed to strong spin--lattice coupling \cite{Hemberger2007,Tachibana2011,kitani2013thermal}, but a general, microscopic theory is lacking. It is therefore interesting to explore the predictions of the microscopic model of spin--lattice coupling, Eq.~(\ref{eq:H}). 
These calculations, which are developed in the Supplemental Material~\footnote[1]\,,\ naturally divide into two parts; (1) an analysis of the different symmetry channels in which the lattice can distort, each with its own associated form of magnetic order; and (2) a characterisation of the spin excitations within each different ordered phase. We find that the dominant magnetic contribution to the thermal expansion comes from the dependence of the $A_1$ (volume) distortion on the magnetization, viz: 

\begin{eqnarray}
\frac{\partial}{\partial T}\frac{\Delta L}{L} = \frac{J \alpha}{K r_0} \frac{8}{3} M \frac{\partial M}{\partial T} \; , 
\label{eq:dL_M}
\end{eqnarray}
where $r_0$ is the equilibrium lattice spacing, and $\alpha$ and $K$ are magnetoelastic couplings defined through Eq.~(\ref{eq:H}). 

NTE will occur when the magnetic contribution, Eq.~(\ref{eq:dL_M}), is both negative and sufficiently large to overcome the usual thermal expansion of the lattice \cite{Hausch1973}. This criterion is easily met in the half-magnetization plateau of CdCr$_2$O$_4$, where $\alpha$ and $M$ are individually large and positive, and the existence of a nearly--localised band of spin excitations at low energies provides a microscopic explanation for the rapid decrease of magnetization with temperature,  $\partial M/\partial T < 0$ \footnote[1]  

.  

 This mechanism finds validation in both experiment \cite{Ueda2005}, where the magnetization is observed to be sharply suppressed by increasing temperature, and in Monte Carlo simulation, as shown in Fig.~S2 and Fig.~S3  \footnote[1]
 
 , and in Ref. \cite{Motome2006}. It is also interesting to note that the NTE must be accompanied by a substantially--enhanced magnetocaloric effect (MCE)

\begin{eqnarray}
\Gamma_{\sf MCE} 
= \frac{\partial T}{\partial H} \Big|_S 
= - \frac{T}{C_H} \frac{\partial M}{\partial T} \Big|_H
\; ,
\label{eq:MCE}
\end{eqnarray}
coming from the same nearly--localised band of excitations \cite{Zhitomirsky2003,Derzhko2006,Schmidt2007}. To the best of our knowledge, this has yet to be measured in experiment. 

In making this analysis, we have assumed in Eq.~(\ref{eq:dL_M}) that $J$ does not vary with temperature: this is a good approximation for changes occurring \emph{within} a given phase, although clearly $J$ can change substantially between phases with different lattice symmetry \cite{Kimura2015}.  We can estimate the fractional change in $J$ with temperature within the plateau phase, from the known dependence of $T_{CW}$ on the Cr-Cr spacing \cite{ueda2008pressure} and the magnitude of the NTE, at $\Delta J / J \approx 1\times10^{-3}$ $K^{-1}$.   We note that ZnCr$_2$Se$_4$  shows a positive Curie--Weiss temperature while ordering antiferromagnetically, and this has been taken to imply that J varies strongly with temperature [50].   However we conclude that this is attribute is not necessary to achieve NTE.

While Eq.~(\ref{eq:dL_M}) has been derived here in the context of the half--magnetization plateau of a Cr spinel, it has a much wider validity, and we would expect NTE to occur in many pyrochlore compounds where the above criteria are met: this is supported by measurements on other Cr spinels. In CdCr$_{2}$S$_4$, NTE is observed to set in below 98~K in the paramagnetic phase, and persists into the ferromagnetic phase \cite{Tachibana2011}: here ${\partial M}/{\partial T}<0$ in both phases. ZnCr$_{2}$Se$_4$ shows NTE below 75~K, but it is suppressed below T$_{N}$ = 21~K \cite{Hemberger2007, Chen2014}, where ${\partial M}/{\partial T}> 0$. We would also predict NTE to occur in the high--field saturated magnetization phase of the oxide spinels \footnote[1]

.

In summary, we made thermal expansion and magnetostriction measurements of the frustrated spinel CdCr$_2$O$_4$, at low temperatures and at magnetic fields up to 30~T. The experimental phase diagram strongly resembles that produced from Monte Carlo simulations of a minimal model of spin--lattice coupling, but diverges in that the plateau phase is more thermally stable than predicted, providing independent verification of the particularly strong spin--lattice coupling in this phase. We also observe NTE in the half--magnetization plateau phase, and show how this can be explained in terms of the same microscopic model. We find the origin of the NTE to be a large, negative temperature--derivative of magnetization, which comes from a band of nearly--localized spin excitations. 

These results are applicable across a broad range of spinel and pyrochlore magnets, and potentially other frustrated magnets. They offer a route to the identification of other new NTE materials, by suggesting that NTE is likely to occur in frustrated magnets where there is a collinear magnetic phase with a flat band. The results also imply a strong link between NTE and an enhanced magnetocaloric effect.

\begin{acknowledgments}

This work was supported by the Dutch funding organization NWO-I and by HFML-RU/FOM, a member of the European Magnetic Field Laboratory (EMFL). R.K. is is supported by the German Science Foundation through Project No. KU 3287/1-1, K.P. by Hungarian NKFIH Grant No. K 124176 and BME - Nanonotechnology and Materials Science FIKP grant of EMMI (BME FIKP-NAT), and N.S. by the Theory of Quantum Matter Unit of the Okinawa Institute of Science and Technology Graduate University. K.P. acknowledges the hospitality of the Theory Quantum Matter Unit, OIST, where part of this work was carried out.

\end{acknowledgments}

\bibliography{CdCr2O4}

\end{document}


\title{Supplemental material for: negative thermal expansion in the plateau state of a magnetically{--}frustrated spinel}

\author{L. Rossi}
\author{A. Bobel}
\author{S. Wiedmann}
\affiliation{High Field Magnet Laboratory (HFML–-EMFL), Radboud University, Nijmegen, Netherlands}
\affiliation{Radboud University, Institute of Molecules and Materials, Nijmegen, Netherlands}
\author{R. K\"uchler}
\affiliation{Max Planck Institute for Chemical Physics of Solids, N{\"o}thnitzer Str. 40, 01187 Dresden, Germany}
\author{Y.~Motome}
\affiliation{Department of Applied Physics, The University of Tokyo, Hongo 7-3-1, Bunkyo-ku, Tokyo 113-8656, Japan}
\author{K. Penc}
\affiliation{Institute for Solid State Physics and Optics, Wigner Research Centre for Physics, Hungarian Academy of Sciences, Budapest, Hungary}
\author{N. Shannon}
\affiliation{Okinawa Institute of Science and Technology Graduate University, 
{Onna--son, Okinawa 904--0495}, Japan}
\affiliation{{Department of Physics, Technische Universit\"at M\"unchen, D-85748 Garching, Germany}}
\author{H. Ueda}
\affiliation{Department of Chemistry, Graduate School of Science, Kyoto University, Kyoto, Japan
}
\author{B. Bryant}
\affiliation{High Field Magnet Laboratory (HFML–-EMFL), Radboud University, Nijmegen, Netherlands}
\affiliation{Radboud University, Institute of Molecules and Materials, Nijmegen, Netherlands}

\maketitle

\section{Zero-field thermal expansion measurements, and effect of varying applied stress}

\begin{figure}[h]
	\begin{centering}
    \includegraphics[scale=1]{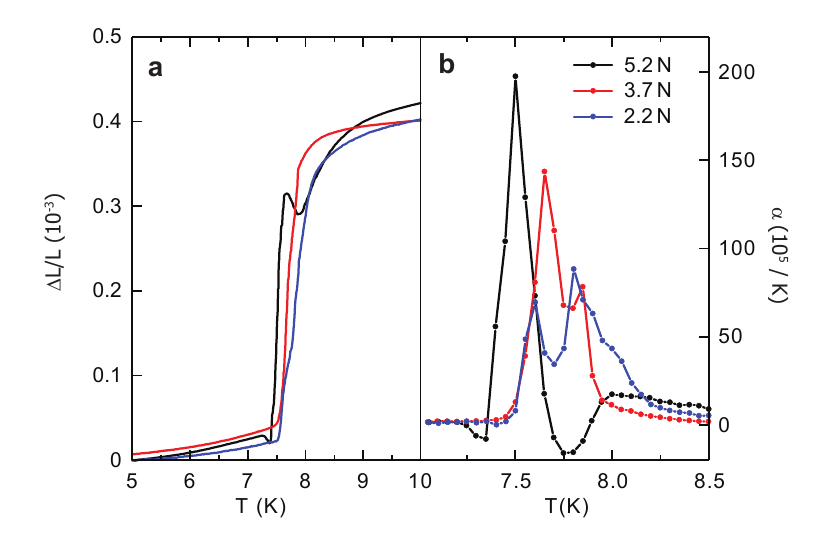}
	\caption{(color online). Magnetostructural transition at T$_N$ in CdCr$_{2}$O$_4$, at zero field. (a) Thermal expansion $\Delta L/L$ and (b) thermal expansion coefficient $\alpha=d(\Delta L/L)/dT$ at zero field from 5~K to 10~K with different uniaxial pressure applied. Increasing the force decreases T$_{N}$ from 7.8~K to 7.5~K.}
	\label{Figure1}
	\end{centering}
\end{figure}

Thermal expansion measurements from 5~K to 10~K at zero field were collected with a capacitive mini--dilatometer in a PPMS [40]. Fig. \ref{Figure1} shows the magnetostructural transition from the cubic \emph{Fd}$\bar{3}$\emph{m}, paramagnetic phase to the tetragonal $I4_1/amd$, antiferromagnetic phase on cooling through T$_{N}$ [31]. In the thermal expansion coefficient $\alpha$ (Fig. \ref{Figure1}(b)) two transitions are seen. Such a splitting has also been seen in specific heat measurements [48], and can be accounted for as an effect of inhomogeneous strain at tetragonal domain boundaries. Fig. \ref{Figure1}(b) also shows the effect of increasing the force applied to the sample from 2.2~N to 5.2~N. Increased uniaxial stress favours one domain orientation, making the sample closer to single-domain and resulting in a sharper transition with a higher $\alpha$. Increasing the applied force also suppresses T$_{N}$ from 7.8~K to 7.5~K. This is an opposite effect to the one seen with the application of hydrostatic pressure, where increasing pressure enhances T$_{N}$ and $\Theta_{CW}$ [34]. We can explain this as due to the applied [111] stress competing with the [001] strain due to the cubic to tetragonal transition. Suppression of antiferromagnetic ordering by uniaxial pressure has also been seen in the chiral magnet MnSi [54].

The data shown in Fig. \ref{Figure1} show a strain of $\approx -4\times10^{-4}$ at T$_N$,  slightly larger than the value of $-1.6\times10^{-4}$ derived from neutron diffraction data [55] and the similar values reported in previous thermal expansion measurements [31]. This discrepancy can be explained by a slight miscut angle between the [111] axis and the direction of applied pressure. Due to the change in lattice symmetry at  T$_N$ the measured strain is highly sensitive to this miscut angle, so it is not possible to give a precise length or volume change at T$_N$. For high field measurements, we polished the samples to have parallel and flat surfaces, to apply a reduced, homogeneous pressure. We applied a force similar to the highest force in Fig. \ref{Figure1}, to attempt to create a single-domain tetragonal state. However, polishing had the effect of further increasing the strain observed at T$_N$, possibly due to an unintended increased miscut or wedge angle between the sample faces.

\newpage
\section{Supplementary data}

\begin{figure}[h]
\includegraphics[width= 0.95 \columnwidth]{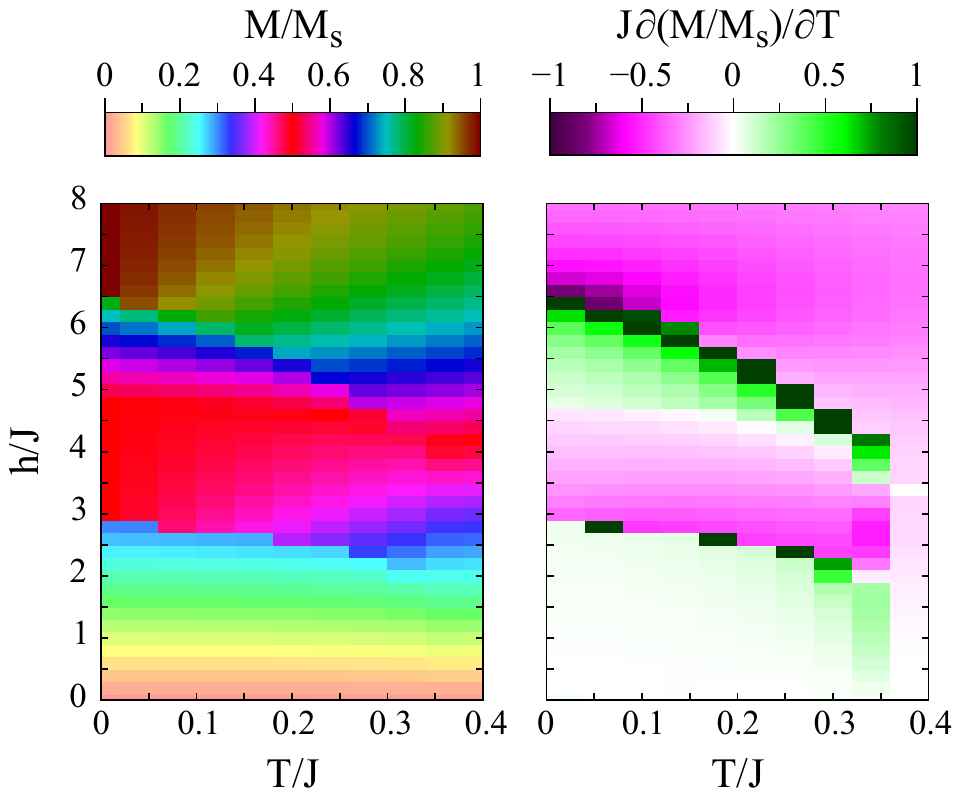}
\caption{The magnetization $M$ (left panel) and its temperature derivative (right panel) as a function of magnetic field and temperature. The magnetization is calculated by Monte-Carlo calculations for $b=0.1$ and 3456 spins, as in [19], and is given as a fraction of the saturation magnetization $M_S$. As per Eq. (\ref{eq:thermal_ex}), negative $M \partial M/\partial T$ leads to negative thermal expansion in the in the lower part of the half magnetization plateau, and above saturation.}
\label{fig:MdMdT}
\end{figure}

\section{The model}

We will derive an expression for the magnetic contribution to the thermal expansion in a pyrochlore lattice, based on a magnetoelastic model [18]. The spin-phonon Hamiltonian we start from is 
\begin{equation}
 \mathcal{H} = \sum_{\langle i,j \rangle} 
  \left[
  J (1- \alpha \rho_{i,j}) \mathbf{S}_i \cdot \mathbf{S}_j
 + \frac{K}{2}  \rho_{i,j}^2 \right] -  \sum_{i} \mathbf{h}\cdot \mathbf{S}_i \,,
\end{equation}
where the summation is over the nearest neighbor bonds on the pyrochlore lattice, $\alpha$ is the spin-lattice coupling, and $K$ the  elastic coupling constant. We will rather use the relative displacements $\delta_{i,j} = \rho_{i,j}/r_0$ (where $r_0$ is the equilibrium distance)
\begin{equation}
 \mathcal{H} = \sum_{\langle i,j \rangle} 
  \left[
  J (1- \tilde \alpha \delta_{i,j}) \mathbf{S}_i \cdot \mathbf{S}_j
 + \frac{\tilde K}{2}  \delta_{i,j}^2 \right] -  \sum_{i} \mathbf{h}\cdot \mathbf{S}_i \,,
 \label{eq:Hmagnetoelastic}
\end{equation}
where 
\begin{align}
  \tilde \alpha &= r_0 \alpha = \frac{r_0}{J(r_0)} \left.\frac{\partial J(r)}{\partial r} \right|_{r=r_0} \,, \\
  \tilde K &= r_0^2 K \,.
\end{align}

Irreducible representations of distances:
\begin{equation}
 \left(
 \begin{array}{c}
  \delta_{{\sf A_1}}  \\
  \delta_{{\sf E},1}  \\
  \delta_{{\sf E},2}  \\
  \delta_{{\sf T_2},1}  \\
  \delta_{{\sf T_2},2}  \\
  \delta_{{\sf T_2},3}  \\
 \end{array}
\right)=
\left(
\begin{array}{cccccc}
\frac{1}{{\sqrt{6}}} & \frac{1}{{\sqrt{6}}} & \frac{1}{{\sqrt{6}}} &
 \frac{1}{{\sqrt{6}}} & \frac{1}{{\sqrt{6}}} & \frac{1}{{\sqrt{6}}} \\
\frac{1}{\sqrt{3}} & \frac{-1}{2\sqrt{3}} & \frac{-1}{2\sqrt{3}} &
 \frac{-1}{2\sqrt{3}} & \frac{-1}{2 \sqrt{3}} & \frac{1}{\sqrt{3}} \\
0 & \frac{1}{2} & -\frac{1}{2}  & -\frac{1}{2}  & \frac{1}{2} & 0 \\
0 & 0 & -\frac{1}{\sqrt{2}} & \frac{1}{\sqrt{2}} & 0 & 0 \\
0 & -\frac{1}{\sqrt{2}} & 0 & 0 & \frac{1}{\sqrt{2}}  & 0 \\
-\frac{1}{\sqrt{2}} & 0 & 0 & 0 & 0 & \frac{1}{\sqrt{2}} \\
\end{array}
\right)
\left(\begin{array}{c} 
 \delta_{1,2} \\
 \delta_{1,3} \\
 \delta_{1,4} \\ 
 \delta_{2,3} \\
 \delta_{2,4} \\
 \delta_{3,4} \\
\end{array}
\right)
\end{equation}
and similarly for the spins, with $\delta_{i,j}$ replaced by $\mathbf{S}_i \cdot \mathbf{S}_j$, and $\delta_{{\sf A_1}}$ by $\Lambda_{{\sf A_1}}$, and so on.

The Hamiltonian of an embedded tetrahedron can then be written as the interactions in the different symmetry channels
\begin{eqnarray}
  \mathcal{H} &=& 2 \sqrt{6} J \Lambda_{\sf{A}} 
   - 2  J 
   \left(
    \tilde\alpha_{\sf{A}} \Lambda_{\sf{A}} \delta_{\sf{A}} 
   +\tilde\alpha_{\sf{E}} \mathbf{\Lambda}_{\sf{E}} \cdot \mathbf{\delta}_{\sf{E}}
   +\tilde\alpha_{\sf{T_2}} \mathbf{\Lambda}_{\sf{T_2}} \mathbf{\delta}_{\sf{T_2}}
   \right) \nonumber \\ 
   && +\left(
   \tilde K_{\sf{A}} \delta_{\sf{A}}^2 
   + \tilde K_{\sf{E}} \mathbf{\delta}_{\sf{E}}^2
   + \tilde K_{\sf{T_2}} \mathbf{\delta}_{\sf{T_2}}^2 \right) \,.
\end{eqnarray}

The energy minima are found for 
\begin{equation}
\delta_{\sf{R}}= \frac{\tilde\alpha_{\sf{R}} J}{\tilde K_{\sf{R}}} \Lambda_{\sf{R}} = \frac{1}{r_0} \frac{\alpha_{\sf{R}} J}{K_{\sf{R}}} \Lambda_{\sf{R}} \,,
\label{eq:deltaR}
\end{equation}
so the energy above becomes
\begin{equation}
  \mathcal{H} = 2 \left( \sqrt{6} J \Lambda_{\sf{A_1}} 
   - b_{\sf{A_1}} \Lambda_{\sf{A_1}}^2
   - b_{\sf{E}} \mathbf{\Lambda}_{\sf{E}}^2
   - b_{\sf{T_2}} \mathbf{\Lambda}_{{\sf T_2}}^2 \right)
   \label{eq:HAET}
\end{equation}
with 
\begin{equation}
  b_{\sf R}=\frac{J \tilde\alpha_{\sf{R}}^2}{2 \tilde K_{\sf{R}}}=\frac{J \alpha_{\sf{R}}^2}{2 K_{\sf{R}}} .
\end{equation}
In Ref. [18] we have considered the case where all the couplings were equal: $b_{\sf{A_1}}=b_{\sf{E}}=b_{\sf{T_2}}$ for simplicity. This also implied that only biquadratic terms of the form $(\mathbf{S}_i\cdot \mathbf{S}_j)^2$ were present in the Hamiltonian. Once we allow different biquadratic coupling strengths, we will get three and four--site terms in the effective Hamiltonian of the form 
$(\mathbf{S}_i\cdot \mathbf{S}_j)(\mathbf{S}_j\cdot \mathbf{S}_k)$ and
$(\mathbf{S}_i\cdot \mathbf{S}_j)(\mathbf{S}_k\cdot \mathbf{S}_l)$, where $i$, $j$, $k$ and $l$ are different site indices. 

The square of the magnetization per site, $\mathbf{M}=(\mathbf{S}_1+\mathbf{S}_2+\mathbf{S}_3+\mathbf{S}_4)/4$, can be expressed via 
$\Lambda_{\sf{A}}$ as
\begin{equation}
  M^2 = \frac{1}{4} + \frac{\sqrt{6}}{8}\Lambda_{\sf{A}} \,.
  \label{eq:MA}
\end{equation}

\section{Theory applied to the plateau phase - space group \#212}

The position of the sites with up and down spins are
\begin{subequations}
\begin{align}
  r_{\downarrow,1} &= a \left( \frac{1}{8}, \frac{1}{8}, \frac{1}{8} \right) \,, \\
  r_{\uparrow,2} &= a \left( \frac{3}{8}, \frac{3}{8} + \xi, \frac{1}{8} + \xi \right) \,,\\
  r_{\uparrow,3} &= a \left(\frac{1}{8} + \xi,\frac{3}{8},\frac{3}{8} + \xi \right) \,,\\
  r_{\uparrow,4} &= a \left(\frac{3}{8} + \xi,\frac{1}{8} + \xi , \frac{3}{8}\right) \,,  
\end{align}
\end{subequations}
where $a$ is the linear size of the cubic unit cell containing 16 sites, and $\xi$ is the displacement of the magnetic ions - we write this for one tetrahedron, the position of the other sites can be concluded. 

From the positions, we can calculate the distances between the sites with antiparallel and parallel spin configurations
\begin{subequations}
\begin{align}
d_{\downarrow\uparrow} &= \frac{a}{2} \sqrt{\frac{1}{2} + 2 \xi + 8 \xi^2} \,,\\
d_{\uparrow\uparrow} &= \frac{a}{2} \sqrt{\frac{1}{2} - 2 \xi + 8 \xi^2} \,,
\end{align}
\end{subequations}
and we can extract $a$ and $\xi$:
\begin{align}
a^2 &= 2 \left(d_{\downarrow\uparrow}^2  + d_{\uparrow\uparrow}^2 + \sqrt{10 d_{\downarrow\uparrow}^2 d_{\uparrow\uparrow}^2 -3 d_{\downarrow\uparrow}^4 - 3 d_{\uparrow\uparrow}^4}
\right) \,,
\\
\xi &= \frac{d_{\downarrow\uparrow}^2 - d_{\uparrow\uparrow}^2}{a^2} \;.
\end{align}
If $d_{\downarrow\uparrow}=d_{\downarrow\uparrow}=d$, we recover the expected result for the F3md,  $a = a_0 = \sqrt{8} d $ and $\xi = 0$.

Now, let us look at the relative length changes: 
\begin{subequations}
\begin{align}
d_{\uparrow\uparrow} \to \frac{a_0}{\sqrt{8}} (1 + \delta_{\uparrow\uparrow}) \,,
\\
d_{\downarrow\uparrow} \to  \frac{a_0}{\sqrt{8}} (1 + \delta_{\downarrow\uparrow}) \,,
\end{align}
\end{subequations}
where we introduced the relative displacements $\delta$.
We get
\begin{align}
\xi &= \frac{\delta_{\downarrow\uparrow}-\delta_{\uparrow\uparrow}}{4} + O(\delta^2)
\\
\frac{a}{a_0} &=  1 + \frac{\delta_{\downarrow\uparrow}+\delta_{\uparrow\uparrow}}{2}  -\frac{3}{8} (\delta_{\downarrow\uparrow}-\delta_{\uparrow\uparrow})^2+O(\delta^3)
\end{align}
so
\begin{align}
\frac{\Delta L}{L} = \frac{\delta_{\downarrow\uparrow}+\delta_{\uparrow\uparrow}}{2}  -\frac{3}{8} (\delta_{\downarrow\uparrow}-\delta_{\uparrow\uparrow})^2+O(\delta^3)
\end{align}

Using the irreps we defined in [18], $\delta_{\mathsf{T_2},1} = \delta_{\mathsf{T_2},2} = \delta_{\mathsf{T_2},3} =  \delta_{\mathsf{T_2}}/\sqrt{3}$ , and 
\begin{subequations}
\begin{align}
\delta_{\uparrow\uparrow} &= \frac{1}{\sqrt{6}} \delta_{\mathsf{A_1}} + \frac{1}{\sqrt{6}} \delta_{\mathsf{T_2}}
, \\ 
\delta_{\downarrow\uparrow} &= \frac{1}{\sqrt{6}} \delta_{\mathsf{A_1}} - \frac{1}{\sqrt{6}} \delta_{\mathsf{T_2}} \,,
\end{align}
\end{subequations}
so that 
\begin{align}
\frac{\Delta L}{L} = \frac{1}{\sqrt{6}}  \delta_{\mathsf{A_1}}   - \frac{1}{4} \delta_{\mathsf{T_2}}^2 +O(\delta^3) \,.
\end{align}
The thermal expansion coefficient is then 
\begin{align}
\frac{\partial}{\partial T} \frac{\Delta L}{L} = \frac{1}{\sqrt{6}}  \frac{\partial \delta_{\mathsf{A_1}}}{\partial T}    - \frac{1}{2} \delta_{\mathsf{T_2}} \frac{\partial \delta_{\mathsf{T_2}}}{\partial T} +O(\delta^3) \;.
\label{eq:thermal_ex_plateau}
\end{align}

\section{Tetragonal case - space group \#141}

The positions of sites in the elongated tetrahedron are
\begin{subequations}
\begin{align}
  r_{R,1} &= \left(\frac{a}{8},\frac{a}{8},\frac{b}{8}\right) \,, \\
  r_{R,2} &= \left(\frac{3 a}{8},\frac{3 a}{8},\frac{b}{8}\right) \,, \\
  r_{L,3} &= \left(\frac{a}{8},\frac{3 a}{8},\frac{3 b}{8}\right) \,, \\
  r_{L,4} &= \left(\frac{3 a}{8},\frac{a}{8},\frac{3 b}{8}\right)  \,, 
\end{align}
\end{subequations}
setting $a=b=a_0$ we recover the undistorted case. The subscript $R$ and $L$ stands for the spins canted  right and left.

The bond lengths are
\begin{subequations}
\begin{align}
 d_{F} &= \frac{a}{\sqrt{8}} \,,\\
d_{C} &= \frac{1}{4}\sqrt{ a^2 + b^2} \,,
\end{align}
\end{subequations}
where the subscript $F$ refers to the bonds connecting parallel spins, i.e. ferro bonds, and $C$ refers to bonds with spins having a canting angle.
From these equations
\begin{subequations}
\begin{align}
 a^2 &= 8 d_F^2 \,,\\
 b^2 &= 16 d_C^2 - 8 d_F^2 \,.
\end{align}
\end{subequations}
Here we again introduce the relative bond stretchings $\delta_F$ and $\delta_C$ as
\begin{subequations}
\begin{align}
d_F = \frac{a_0}{\sqrt{8}} (1 + \delta_F) \,,
\\
d_C =  \frac{a_0}{\sqrt{8}} (1 + \delta_C) \,.
\end{align}
\end{subequations}
Then
\begin{align}
 \frac{\Delta a}{a_0} &= \delta_F,\\
\frac{\Delta b}{a_0} &= 2\delta_C - \delta_F - (\delta_C - \delta_F)^2 + O(\delta^3), \\
\frac{\Delta V}{V_0} &= (2 \delta_C + \delta_F) - (\delta_C^2 - 6 \delta_C \delta_F + 2 \delta_F^2)  + O(\delta^3) \,,
\end{align}
and the $\Delta L/L$ calculated from the volume change is
\begin{align}
\frac{\Delta L}{L} 
&= \frac{V^{1/3}-V_0^{1/3}}{V_0^{1/3}} \nonumber\\
&= \frac{1}{3} (2 \delta_C + \delta_F) - 
 \frac{7}{9} (\delta_C - \delta_F)^2  +  O(\delta^3) \nonumber\\
 &= \frac{1}{\sqrt{6}} \delta_{\sf{A_1}} - 
 \frac{7}{12} \delta_{\sf{E}}^2+ O(\delta^3) \,.
 \label{eq:thermal_ex_tetra})]
\end{align}
We used above that 
\begin{subequations}
\begin{align}
  \delta_{\sf{A_1}}& = \frac{2}{\sqrt{6}} (\delta_F + 2 \delta_C ) \,,\\
  \delta_{\sf{E},1}& = \frac{2}{\sqrt{3}} (\delta_F - \delta_C) \,, \\
  \delta_{\sf{E},2}& = 0 \,, \\
  \delta_{\sf{T},j}& = 0  \,.
\end{align}
\end{subequations}

\begin{table}[tb]
\caption{\label{table:DLoL} Results for magnetostriction in the collinear half--magnetization plateau and tetragonal canted phases, derived from the microscopic model of spin--lattice coupling, Eq.~(\ref{eq:Hmagnetoelastic}).
The expression for the thermal expansion coefficient in the two phases is identical at leading order.
}
\begin{ruledtabular}
   \begin{tabular}{ccc}
 phase: & plateau (${\sf T_2}$)  & tetragonal (${\sf E}$) \\
space group: & \#212 & \#141 \\
\hline 
$\frac{\Delta L}{L}$ & $\frac{1}{\sqrt{6}}  \delta_{\mathsf{A_1}}   - \frac{1}{4} \delta_{\mathsf{T_2}}^2 +O(\delta^3) $&$ \frac{1}{\sqrt{6}} \delta_{\sf{A_1}} - 
 \frac{7}{12} \delta_{\sf{E}}^2+ O(\delta^3) $\\
$\frac{\Delta h}{h}$ & not applicable &$\frac{1}{\sqrt{6}} \delta_{\sf{A_1}} - 
 \frac{5}{4} \delta_{\sf{E}}^2+ O(\delta^3) $\\
$\frac{\partial}{\partial T}\frac{\Delta L}{L}$ & $\frac{\tilde\alpha_{\sf{A_1}} J}{\tilde K_{\sf{A_1}}} \frac{8}{3} M \frac{\partial M}{\partial T} + \cdots$ & $\frac{\tilde\alpha_{\sf{A_1}} J}{\tilde K_{\sf{A_1}}} \frac{8}{3} M \frac{\partial M}{\partial T} + \cdots$ \\
   \end{tabular}
\end{ruledtabular}
\end{table}

However, we measure $\Delta L/L$ along the original [111] distances. We assume it will be the direction which is orthogonal to a face of the elongated tetrahedron.
The volume of a single tetrahedron is
\begin{equation}
 V_{\text{tetra}} = \frac{1}{4^3}\frac{a^2b}{3} \,,
\end{equation}
the area of the triangle (the face of the tetrahedron) is  
\begin{equation}
A_{\text{face}} = \frac{1}{4^2}\frac{a \sqrt{a^2+2 b^2}}{2}  \,,
\end{equation}
so the height is 
\begin{equation}
h_\perp = \frac{1}{4} \frac{2 a b}{\sqrt{a^2+2 b^2}} \,.
\end{equation}
We can express this using the bond lengths as
\begin{equation}
h_\perp = \frac{d_F \sqrt{4 d_C^2 - 2 d_F^2}}{\sqrt{4 d_C^2 - d_F^2}} \,.
\end{equation}
The undistorted $h_{\perp,0} = \sqrt{2/3} d_0$, and for the relative height change we get
\begin{equation}
\frac{\Delta h}{h} = \frac{1}{3} (\delta_F + 2 \delta_C) - 
 \frac{5}{3} (\delta_F - \delta_C )^2 + O(\delta^3) \,.
\end{equation}
The leading term is the same as the one calculated from the volume change, the difference is in the prefactor of the second order term. 
\begin{equation}
\frac{\Delta h}{h} = \frac{1}{\sqrt{6}} \delta_{\sf{A_1}} - 
 \frac{5}{4} \delta_{\sf{E}}^2+ O(\delta^3) \,.
 \label{eq:thermal_ex_tetra2}
\end{equation}

In both the plateau [Eq. (\ref{eq:thermal_ex_plateau})] and the tetragonal case [Eq. (\ref{eq:thermal_ex_tetra2})] the displacements are linear with the $\delta_{\sf{A_1}}$, so we can assume that the most important part is coming from the dependence of the $\delta_{\sf{A_1}}$ on magnetization. Therefore, we can neglect the quadratic and higher order terms, making the expression for both plateau and tetragonal phases identical. Thus, employing Eqs. (\ref{eq:deltaR}) and (\ref{eq:MA}), we can write a general expression for the magnetic component of the thermal expansion in a pyrochlore lattice:
\begin{align}
\frac{\partial}{\partial T}\frac{\Delta L}{L} &= \frac{1}{\sqrt{6}} \frac{\partial \delta_{\sf{A_1}}}{\partial T}  + \cdots \nonumber\\
  & = \frac{\tilde\alpha_{\sf{A_1}} J}{\tilde K_{\sf{A_1}}} \frac{1}{\sqrt{6}}\frac{\partial \Lambda_{\sf{A_1}}}{\partial T}  + \cdots \nonumber\\
  & = \frac{\tilde\alpha_{\sf{A_1}} J}{\tilde K_{\sf{A_1}}} \frac{8}{3} M \frac{\partial M}{\partial T}  + \cdots
\label{eq:thermal_ex}
\end{align}

Negative thermal expansion will occur when $\Delta L/ L$ is negative and sufficiently large to overcome the normal lattice positive thermal expansion, i.e. when the spin-lattice coupling $\alpha_{\sf{A_1}}$ is large, and the product of M and ${\partial M}/{\partial T}$ is large and negative.

\title{NTE in CdCr2O4 - low-T expansion}

\author{noted by Karlo Penc}
\affiliation{Institute for Solid State Physics and Optics, Wigner Research Centre for Physics, Hungarian Academy of Sciences, H-1525 Budapest, P.O.B. 49, Hungary}

\date{\today ~at \currenttime}

\maketitle

 \section{$\partial M/\partial T$ from the low temperature expansion}

Here we consider the simplest 4-sublattice ordered plateau state, stabilized by the ferromagnetic $J_3$ exchanges. In the low-temperature expansion we integrate out the quadratic fluctuations around the zero-temperature state. For the plateau state, the fluctuations of the spins can be parameterized by the $\xi_{\mathbf{r},i}$ and $\zeta_{\mathbf{r},i}$ real variables for each of the four sublattices $i=0,1,2,3$ in the unit cell at position $\mathbf{r}$, so that 
\begin{align} 
s_{\mathbf{r},1} &= \left( \xi_{\mathbf{r},1}, \zeta_{\mathbf{r},1}, 1-\frac{ \xi_{\mathbf{r},1}^2+ \zeta_{\mathbf{r},1}^2}{2}\right)\;,
\nonumber\\
s_{\mathbf{r},2} &= \left( \xi_{\mathbf{r},2}, \zeta_{\mathbf{r},2}, 1-\frac{ \xi_{\mathbf{r},2}^2+ \zeta_{\mathbf{r},2}^2}{2}\right)\;,
\nonumber\\
s_{\mathbf{r},3} &= \left( \xi_{\mathbf{r},3}, \zeta_{\mathbf{r},3}, 1-\frac{ \xi_{\mathbf{r},3}^2 + \zeta_{\mathbf{r},3}^2}{2}\right)\;,
\nonumber\\
s_{\mathbf{r},4} &= \left( \xi_{\mathbf{r},4}, \zeta_{\mathbf{r},4},-1+\frac{ \xi_{\mathbf{r},4}^2 + \zeta_{\mathbf{r},4}^2}{2}\right)\;.
\end{align}
The energy expression is a quadratic form of the $\xi$ and $\zeta$ variables, which can be conveniently written using Fourier transformation,
\begin{equation}
  \mathcal{E} = \mathcal{E}_0 
  + \frac{1}{2} \sum_{\mathbf{k}\in BZ} \left[ \bm{\xi}_{-\mathbf{k}}\cdot \Lambda(\mathbf{k})\cdot \bm{\xi}_{\mathbf{k}} + \bm{\zeta}_{-\mathbf{k}}\cdot \Lambda(\mathbf{k})\cdot \bm{\zeta}_{\mathbf{k}} \right] \;,
  \label{eq:LT.E_Plateaux}
\end{equation}
where
\begin{equation}
 \mathcal{E}_0 = - \left( \frac{h}{2} +3 b J \right) N 
\end{equation}
is the energy of the ordered classical configuration, the
\begin{align}
\bm{\xi}_{\mathbf{k}} &= \left( \xi_{\mathbf{k},1},\xi_{\mathbf{k},2},\xi_{\mathbf{k},3},\xi_{\mathbf{k},4} \right) \,,\nonumber \\
\bm{\zeta}_{\mathbf{k}} &= \left( \zeta_{\mathbf{k},1},\zeta_{\mathbf{k},2},\zeta_{\mathbf{k},3},\zeta_{\mathbf{k},4} \right) \,, 
\end{align}
are the Fourier transforms of the $\xi_{\mathbf{r},i}$ and $\zeta_{\mathbf{r},i}$, and
\begin{widetext}
\begin{equation}
\Lambda(\mathbf{k}) = \left(
\begin{array}{cccc}
 h-2 (1-6 b) J & 2 (1-2 b) J \cos\frac{k_y-k_z}{2} & 2 (1-2 b) J \cos\frac{k_x-k_z}{2} & 2 (1+2 b) J \cos\frac{k_x+k_y}{2} \\
 2 (1-2 b) J \cos\frac{k_y-k_z}{2} & h-2 (1-6 b) J & 2 (1-2 b) J \cos\frac{k_x-k_y}{2} & 2 (1+2 b) J \cos\frac{k_x+k_z}{2} \\
 2 (1-2 b) J \cos\frac{k_x-k_z}{2} & 2 (1-2 b) J \cos\frac{k_x-k_y}{2} & h-2 (1-6 b) J & 2 (1+2 b) J \cos\frac{k_y+k_z}{2} \\
 2 (1+2 b) J \cos\frac{k_x+k_y}{2} & 2 (1+2 b) J \cos\frac{k_x+k_z}{2} & 2 (1+2 b) J \cos\frac{k_y+k_z}{2} & -h+6 (1+2 b) J \\
\end{array}
\right)
-4 J_3 \gamma_3(\mathbf{k}) \mathbb{1}
\end{equation}
\end{widetext}
is a real matrix describing the energy of the fluctuations, the $\mathbb{1}$ denotes the $4\times 4$ identity matrix, and 
\begin{equation}
\gamma_3(\mathbf{k})= 3-\cos k_x \cos k_y -\cos k_x \cos k_z -\cos k_y \cos k_z \;.
\end{equation}
We assume that there are $N$ spins in the system, so that the number of the unit cells is $N/4$, and this is also the number of the $\mathbf{k}$ points in the Brillouin zone of the effective fcc lattice formed by the four-site unit cells (tetrahedra).


The associated partition function can be calculated as
\begin{align}
\mathcal{Z}^{\sf LT} 
&= 
  \left( \frac{1}{\sqrt{2 \pi}} \right)^{2N} 
  \prod_{\mathbf{k}\in \text{BZ}} 
  \prod_{i=1}^{4} \int d \xi_{\mathbf{k},i} \int d \zeta_{\mathbf{k},i}
e^{-\mathcal{E}/T}
\nonumber \\
 & =  e^{-\mathcal{E}_0/T}  \prod_{\mathbf{k}\in \text{BZ}} 
  \frac{T^4}{\det  \Lambda(\mathbf{k})}  
  \label{eq:partitionfunction}
\end{align}
It follows that, for $T \to 0$, the free energy of the system is given by
\begin{align}
\mathcal{F}^{\sf LT} &=
   {\mathcal E}_0 
   + T \sum_{\mathbf{k} \in \text{BZ}}  \ln \det \Lambda(\mathbf{k}) 
   - N T \ln T 
   + {\mathcal O}(T^2)\,,  \nonumber \\
   \label{eq:FlowT}
\end{align}
where the ${\mathcal O}(T^2)$ corrections arise from the higher order terms neglected in the expression for the energy, Eq.~ (\ref{eq:LT.E_Plateaux}).
The magnetization per site is 
\begin{equation}
  M = -\frac{1}{N} \frac{\partial \mathcal{F}^{\sf LT}}{\partial h} \;,
\end{equation}
and, consequently, the temperature derivative of the magnetization is 
\begin{align}
  \frac{\partial M}{\partial T} &= - \frac{1}{N} \sum_{\mathbf{k} \in \text{BZ}}
  \frac{1}{\det \Lambda(\mathbf{k})} \frac{\partial\det \Lambda(\mathbf{k})}{\partial h} 
  \nonumber \\
    &= -\frac{1}{N} \sum_{\mathbf{k} \in \text{BZ}} \sum_{j=1}^{4}
   \frac{1}{\lambda_j(\mathbf{k})} \frac{\partial\lambda_j(\mathbf{k})}{\partial h} \;,
\end{align}
where $\lambda_{\nu}(\mathbf{k})$, with $\nu=1,\dots,4$, are the four eigenvalues of the $\Lambda(\mathbf{k}) $ matrix, and $\det \Lambda(\mathbf{k}) = \prod_{j=1}^4 \lambda_j(\mathbf{k})$.
These four eigenvalues form bands in the $\mathbf{k}$-space, the first one is 
\begin{equation}
\lambda_1(\mathbf{k}) = h -4 J +16 b J - 4 J_3 \gamma_3(\mathbf{k}) \;,
 \label{eq:lambda1}
\end{equation}
the dispersion is provided by the $J_3$ ferromagnetic coupling -- in the absence of $J_3$ the band is `flat' and describes localized modes which will become important later on. The remaining three eigenvalues are solution of a cubic equation, and can be easily computed numerically.

\begin{figure}
\includegraphics[width= 0.98 \columnwidth]{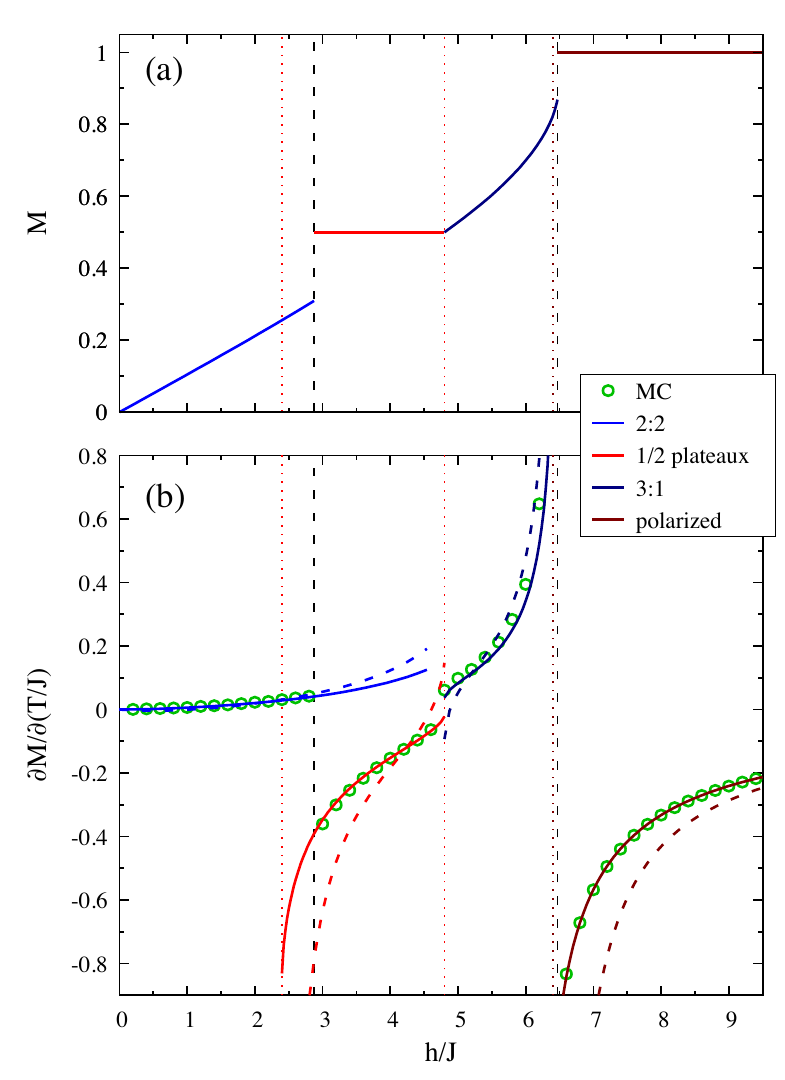}
\caption{(a) The magnetization $M$ and (b) the temperature derivative of the magnetization, $J \partial M/\partial T$, as calculated from the low-temperature expansion (solid curves) and MC calculation (open circles) for $b=0.1$ and $J_3/J=-0.05$.  The dotted vertical lines show the local instabilities of the $M=1/2$ plateau at $h/J=2.4$ and $h/J=4.8$, and of the $M=1$ polarized phase at $h=6.4$. For $J_3=0$ (dashed curves), the corresponding $J \partial M/\partial T$  diverge at lower instability fields. The vertical dashed lines show the first order transition at the lower edge of the plateau and at the lower edge of the polarized state.}
\label{fig:lowT_vs_MC}
\end{figure}

We have derived the corresponding expressions in the other phases as well. Our findings are summarized in 
Fig.~\ref{fig:lowT_vs_MC}. The calculated $\frac{\partial M}{\partial T}$ follows the MC results, thus reinforcing the reliability of the approach. More importantly, we also observe that the  $\frac{\partial M}{\partial T}$ in the plateau ($M=1/2$) and in the polarized ($M=1$) phase actually diverges as the lower critical fields are approached, only to be cut off by the finite value of the $J_3$ and by the first order phase transition field. This divergency is in fact associated to the gap-closing of the $\lambda_1(\mathbf{k})$, given by Eq.~(\ref{eq:lambda1}). Namely, the
 contribution from the narrow band to the temperature derivative of magnetization is
\begin{align}
 \frac{\partial M^{\text{NB}}}{\partial T}  
  &= -\frac{1}{N} \sum_{\mathbf{k} \in \text{BZ}} 
   \frac{1}{\lambda_1(\mathbf{k})} \frac{\partial\lambda_1(\mathbf{k})}{\partial h} 
   \nonumber\\
   &= -\frac{1}{N} \sum_{\mathbf{k} \in \text{BZ}} 
   \frac{1}{h - (4 -16 b)  J - 4 J_3 \gamma_3(\mathbf{k})} \;. 
\end{align}
At the lower critical field, $h_c = (4 - 16 b)  J$, the derivative diverges with the narrowing bandwidth
\begin{align}
 \left.   \frac{\partial M^{\text{NB}}}{\partial T}  \right|_{h=h_c}
    &= 
  \frac{1}{N} \sum_{\mathbf{k} \in \text{BZ}} 
   \frac{1}{4 J_3 \gamma_3(\mathbf{k})} 
   \nonumber\\  
   &\approx \frac{0.4482}{16 J_3 } = \frac{0.028014}{J_3 } \;,
\end{align}
where we used that 
\begin{equation}
 \int\limits_{-\pi}^{\pi}\int\limits_{-\pi}^{\pi} \int\limits_{-\pi}^{\pi} \frac{d^3 \mathbf{k}}{(2\pi)^3} \frac{1}{\gamma_3(\mathbf{k})}
  = \frac{\sqrt{3}K\left(\frac{\sqrt{3}-1}{\sqrt{8}} \right)}{\pi ^2} \approx 0.4482 \;.
\end{equation}
In the absence of $J_3$, the $\lambda_1(\mathbf{k})$ band is dispersionless (flat), and we get
\begin{align}
 \left. \frac{\partial M^{\text{NB}}}{\partial T} \right|_{J_3=0}   
   &= -\frac{1}{N} \sum_{\mathbf{k} \in \text{BZ}}  
   \frac{1}{h - (4 -16 b)  J } 
   \nonumber\\ 
     &= -\frac{1}{4} 
   \frac{1}{h - h_c } \;,
\end{align}
i.e. it diverges as the $h \rightarrow h_c$ from above. 

Thus we can associate the large NTE with the localized modes of collinear states. 

The divergency is also present at the $h_c^{\text{FM}}= (8-16b)J$ critical field where the all the spins align with the external magnetic field, i.e. $M=1$. 
 In this case the 
\begin{widetext}
\begin{equation}
\Lambda(\mathbf{k})=\left(
\begin{array}{cccc}
 h-6 (1-2 b)  J  & 2 (1-2 b)  J  \cos\frac{k_y-k_z}{2} & 2 (1-2 b)  J  \cos\frac{k_x-k_z}{2} & 2 (1-2 b)  J  \cos\frac{k_x+k_y}{2} \\
 2 (1-2 b)  J  \cos\frac{k_y-k_z}{2} & h-6 (1-2 b)  J  & 2 (1-2 b)  J  \cos\frac{k_x-k_y}{2} & 2 (1-2 b)  J  \cos\frac{k_x+k_z}{2} \\
 2 (1-2 b)  J  \cos\frac{k_x-k_z}{2} & 2 (1-2 b)  J  \cos\frac{k_x-k_y}{2} & h-6 (1-2 b)  J  & 2 (1-2 b)  J  \cos\frac{k_y+k_z}{2} \\
 2 (1-2 b)  J  \cos\frac{k_x+k_y}{2} & 2 (1-2 b)  J  \cos\frac{k_x+k_z}{2} & 2 (1-2 b)  J  \cos\frac{k_y+k_z}{2} & h-6 (1-2 b)  J  \\
\end{array}
\right)
-4 J_3 \gamma_3(\mathbf{k}) \mathbb{1} \;,
\end{equation}
\end{widetext}
and there are two narrow bands with 
\begin{equation}
 \lambda(\mathbf{k})_{1,2} = h - 8 J +16 b  J - 4 J_3 \gamma_3(\mathbf{k}).
\end{equation}
The doubling of the flat modes with respect to the $M=1/2$ plateau case is reflected in the roughly twice as large $\partial M/\partial T$, clearly seen in Fig.~\ref{fig:lowT_vs_MC}(b).